# Impacts of thermal aging and associated heat losses on the performance of a Pyromark 2500-coated concentrated solar power central receiver

Katie Bezdjian[1,2] and Mathieu Francoeur[3,*]

[1]Instituto de Energía Solar, Universidad Politécnica de Madrid, 28040 Madrid, Spain

[2]Department of Mechanical Engineering, University of Utah, Salt Lake City, UT 84112, USA

[3]Department of Mechanical Engineering, McGill University, Montréal, QC H3A 0C3, Canada

**Abstract**

Pyromark 2500 is a widely used absorber coating for concentrated solar power central receiver systems due to its high absorptivity, ease in application, and relatively low cost. Pyromark's performance is quantified by its figure of merit (FOM), which relates the coating's heat losses to its solar-to-thermal conversion efficiency. After long-term exposure to high temperatures (>750°C) and irradiance levels, Pyromark's absorptivity and FOM decrease. The aim of this research is to evaluate changes in Pyromark's absorptivity, heat losses, and FOM as a function of thermal aging. This work also compares the most common FOM expression, which neglects convection losses, to an FOM that includes all heat losses experienced by a central receiver. Isothermal aging experiments are conducted on Pyromark-coated Inconel 600 substrates at 750°C. The spectral, hemispherical absorptivity of the samples is measured at room temperature with a spectrophotometer and input into a finite element analysis model that includes radiation and convection boundary conditions. The heat flux and temperature output by the model are used to determine the heat losses and FOM of the Pyromark samples. After 151 h of thermal aging, the sample with the thinnest Pyromark coat maintains the most stable total, hemispherical absorptivity.

---

[*] Corresponding author. Email address: mathieu.francoeur@mcgill.ca

Conversely, the total, hemispherical absorptivity of the sample with the thickest Pyromark coat drops by a maximum of 1.73%, and the corresponding maximum drop in FOM is 1.90% when windy conditions (which are expected around central receivers) are assumed. In windy conditions, convection losses constitute between 21% and 24% of the samples' total heat loss; thus, the most common FOM expression in the literature overestimates the samples' FOM by ~4.40%. An analysis of the samples' heat losses indicates that reflection losses exceed emission losses when the absorptivity declines significantly.

# 1. Introduction

Concentrated solar power (CSP) systems have emerged as a viable alternative to fossil fuels due to their utility scale, lower $CO_2$ emissions [1], and option for thermal storage, which increases electricity dispatchability [2]. Central receiver systems are especially promising because their operating temperatures can reach 1000°C, which is higher than other CSP families [3]; this is advantageous for thermal-to-electrical conversion per Carnot's efficiency [4]. Higher operating temperatures also reduce the cost of thermal storage [5].

The receiver in central receiver systems governs the solar-to-thermal energy conversion efficiency and is thusly a key component of these systems [6]. Optimized receiver designs correspond to higher plant efficiencies, improved coupling with thermal energy storage, and cost reductions [7]. Absorber coatings are a common method of enhancing a central receiver's solar-to-thermal energy conversion. However, central receivers are subjected to extreme temperatures and massive irradiance fluctuations between day and night, as well as rapid oscillations in solar flux due to shading from fast-moving clouds [8]. These harsh conditions accelerate the physical and chemical mechanisms that thwart a coating's solar-to-thermal efficiency in a process called thermal aging [9]. Supercritical $CO_2$ (s$CO_2$) Brayton cycles are considered the best-fit power cycle for enhancing the solar-to-thermal conversion efficiency of next generation CSP systems, and these cycles require turbine inlet temperatures ≥ 700°C [10], [11], which is roughly 150°C higher than the current subcritical steam Rankine cycles [10], [12]. Therefore, absorber coatings must have the resiliency to withstand elevated temperatures and thousands of heating and cooling cycles. Other important criteria for next-generation absorber materials are stability in air, applicability at large scales on-site, and cost-effectiveness [13].

The solar-to-thermal conversion efficiency of a receiver coating is characterized by its figure of merit (FOM). A coating's FOM is usually defined as the ratio of the net radiative energy absorbed and retained by the coating to the incident irradiation [14]: $\text{FOM} = (\alpha C q''_{sol} - \varepsilon \sigma T^4)/C q''_{sol}$, where $\alpha$ is the coating's total, hemispherical absorptivity to solar irradiation, $C$ is the concentration ratio of the central receiver system, $q''_{sol}$ is the incident solar flux, $\varepsilon$ is the coating's total, hemispherical emissivity at the surface temperature $T$, and $\sigma$ is the Stefan-Boltzmann constant [$5.67 \times 10^{-8}$ W/(m$^2$·K$^4$)]. Ideal coatings demonstrate high absorptivity in the solar spectral band (visible and near-infrared wavelengths) and low emissivity at longer infrared wavelengths emitted at the surface temperature $T$ [15]. Solar selective absorber coatings can achieve this goal. A number of promising solar selective absorber coatings have been fabricated at the laboratory scale [6], [7], [16]–[24], but these coatings are not yet commercially viable because they are difficult to apply on large parts and reparation techniques have not been developed [2]. Solar selective absorber coatings may be unfavorable for next generation central receiver systems because spectral selectivity is difficult to accomplish at temperatures beyond 700°C due to overlap between the solar spectrum and the coating's emission spectrum [25]. Also, solar irradiation is much larger than thermal emission for concentration ratios $C$ greater than 1000 [6], which are typical for central receiver systems [1], [3], [12], [26], [27]. For these systems, maximizing a coating's absorptivity to solar irradiation is more advantageous than minimizing its emissivity, and this can be achieved with a simpler non-selective solar absorber [21], [25], [28]. Increasing a coating's absorptivity to solar irradiation is also more cost-effective than minimizing its emissivity [29], [30].

Pyromark 2500, which is a high-temperature silicone-based paint manufactured by Tempil [31], is the most popular non-selective absorber coating in central receiver systems [32], [33] due to its manufacturability, relatively low cost, and ease in application [29], including an established and feasible method of repainting in the field [34]. It is comprised of silicate binder with transition metal oxides [2], [34], and its manufacturer claims Pyromark can withstand temperatures up to 1093°C [31]. Pyromark's radiative properties resemble that of a gray and diffuse surface; it is characterized by a high absorptivity within the solar spectrum and a high emissivity in the infrared spectral band [35]. Prior measurements of pristine Pyromark's (i.e., Pyromark that has been cured but has not been subjected to additional aging) radiative properties for wavelengths between 0.3 µm and 4 µm indicated that its spectral, directional absorptivity was independent of temperature up to 600°C and independent of incidence angle between 0° and 40°. A slight decrease in spectral, directional absorptivity occurred at an incidence angle of 60°, and a larger drop occurred at 80° [36]. Nevertheless, it is reasonable to assume that pristine Pyromark is diffuse within the solar spectrum. Aged Pyromark, however, has exhibited significant changes in spectral, directional absorptivity at 40° and 60° [37]. The total, hemispherical absorptivity to solar irradiation of pristine Pyromark is typically between 0.96 and 0.97 [8], [9], [22], [36]–[40]. Pyromark's emissivity is its lesser-studied radiative property. Large discrepancies exist between published emissivity values [22], [36]–[38], [41]–[43], which were attributed to extrinsic factors that may affect coatings' emissivity, such as non-homogeneous surfaces and different curing procedures [22]. In [36], the total, hemispherical emissivity of Pyromark was reported as ~0.8–0.9 for temperatures ranging from 100°C to 1000°C, while total, hemispherical emissivity values of ~0.71–0.76 were reported for temperatures between 200°C and 800°C in [22]. Despite these discrepancies, [22], [36], [38], and [43] reported that Pyromark's total, hemispherical emissivity was relatively independent of temperature beyond 500°C.

Pyromark's durability and radiative properties depend on fabrication parameters such as deposition method, curing temperature, coating thickness [32], [38], [40], [44], and the substrate upon which the coating is applied [37]. In [45], Pyromark samples cured at maximum temperatures of 800°C and 900°C were characterized by lower, more variable total, hemispherical absorptivity values than samples cured at a maximum temperature of 700°C. The Pyromark samples cured at 700°C had a porous structure without well-defined crystal shapes, while the samples cured at 900°C had larger, well-defined crystal structures on the outermost surface. These large crystals reduced the samples' external surface porosity and, consequently, their absorptivity. The total, hemispherical absorptivity of samples cured at 800°C and 900°C were consistent with absorptivity values reported in the literature for aged Pyromark [38], [40], signifying that aging occurred during curing. Coat thickness has also influenced Pyromark's total, hemispherical absorptivity. Coventry and Burge [40] found that thinner Pyromark coatings maintained their total, hemispherical absorptivity better than thicker coatings when aged isothermally at 750°C, but the inverse was true for aging at 850°C. Furthermore, thick Pyromark coatings have delaminated frequently during curing [40], [44], as have samples with four or more layers of Pyromark [44].

The aging process of absorber coatings is accelerated by high irradiances and temperatures [35], and these aging factors are unavoidable in central receiver systems. Limited data on Pyromark's aging behavior has been recorded from actual central receivers [42], and this lack of information has necessitated aging experiments. Two main categories of thermal aging experiments have been performed: fixed aging and cyclic aging.

Fixed aging experiments include isothermal aging and constant irradiance aging. During isothermal aging experiments, Pyromark samples are heated to high temperatures either slowly or rapidly, then held at the target temperature for an extended period. Isothermal aging is believed to accelerate Pyromark's degradation by increasing oxidation rates and thermal stresses in Pyromark-coated substrates [45]. Pyromark has repeatedly demonstrated stability when aged isothermally at temperatures below 700°C [13], [38], [40], [44], but degradation was reported for multiple substrates when the paint was aged isothermally at temperatures approximately equal to or higher than 750°C [13], [21], [36]–[38], [40], [45]. When Pyromark-coated Inconel 625 samples were aged isothermally at 700°C, secondary phases that corresponded to a slight decline in Pyromark's absorptivity began to appear, but only after extensive aging. However, these secondary phases appeared much sooner when Pyromark was aged at 800°C [13]. A similar pattern was observed in [37], which was attributed to increased diffusion and reaction rates at higher temperatures. Constant irradiance aging experiments subject Pyromark samples to various fixed irradiance levels for a predetermined exposure period [9], [39]. Boubault *et al.* [9], [39] performed constant irradiance aging experiments and showed that Pyromark's total absorptivity at a normal angle of incidence decreased with increasing exposition time at nearly every irradiance level.

Cyclic aging mimics the extreme variation in solar irradiance to which central receivers are exposed by rapidly heating and cooling a sample [19], [23], [45], or by subjecting the sample to fluctuating irradiance levels [8], [42], [44], [46]. Boubault *et al.* [46] conducted cyclic aging experiments by exposing Pyromark-coated Inconel 625 samples to square-shaped, periodic irradiance cycles with different mean irradiances, amplitudes, periods, and exposure times. Pyromark's absorptivity and thermophysical properties decreased with higher mean irradiance levels, exposure times, and periods; the FOM of the samples also decreased with increasing irradiance levels. Isothermal aging and cyclic aging have been compared in [44] and [45]. In [44], various substrates were coated with Pyromark and aged isothermally at 700°C. Afterward, ten intact coatings were aged for 5–16 cycles using an average concentration ratio of 600. Many of the samples degraded quickly when exposed to cyclic aging, despite their stability after isothermal aging. This degradation was attributed to intensified thermal expansion differences between the coating and substrate due to the faster heating and cooling rates [44]. Three aging patterns were compared in [45]: isothermal, rapid cycle, and cycle-and-hold, which combined isothermal and rapid cycle aging. Changes in absorptivity resulting from cycle-and-hold aging were not a linear combination of the evolutions in absorptivity due to isothermal and rapid cycle aging. This finding suggested that isothermal and rapid cycle aging have non-linear effects on coating degradation, and neither aging scheme typified Pyromark's degradation when conducted separately.

Pyromark's decrease in FOM as a function of thermal aging is of great interest because the FOM provides a more detailed description of a coating's performance than total, hemispherical absorptivity alone. The typical FOM definition only considers losses due to reflection and emission, but convection losses also impact the energy absorbed and retained by a coating and, consequently, its FOM. Conduction losses from the front to back of the receiver panel are minor compared to other thermal losses [47]–[49]. Past research characterizing convection losses from central receivers [49]–[54] has emphasized the complex flow field surrounding receivers resulting from their massive size [50], [53] and the superposition effects of free and forced convection on the receiver [16], [49], [50], [53]. Uncertainty in convection losses has been declared the highest of all thermal losses, especially at scales as large as central receivers [32]. Still, convection losses have been identified as a major source of thermal losses for central receivers [15]. Convection losses were included in some evaluations of Pyromark's FOM as a function of thermal aging [42], [46], [55], but the vast majority only featured emission and reflection losses. Comparisons of all heat losses that impact a central receiver's performance have been published [7], [47]–[49], [52], [54], [56], [57], but the effects of Pyromark's degradation on heat losses were not considered.

Certain aspects of Pyromark's performance on central receivers (e.g., radiative properties, degradation due to thermal aging) have been studied. However, the effect of Pyromark aging on a central receiver's heat losses is absent from the literature, and FOM analyses that include all heat losses are limited. The objective of this work is to fill this knowledge gap by analyzing the impacts of thermal aging and associated heat losses on the FOM of a Pyromark-coated central receiver. Here, Pyromark samples are subjected to thermal aging experiments to evaluate the paint's absorptivity as a function of high-temperature exposure (Section 2). The measured absorptivity values are then used as parameters in a finite element analysis (FEA) model that quantifies the impact of thermal aging on heat losses and FOM by computing the useable heat flux from a Pyromark-coated central receiver panel (Section 3).

## 2. Thermal aging experiments

### 2.1. Sample preparation and oven treatments

In this work, Pyromark was deposited on four 3.81 cm × 3.81 cm × 0.05 cm Inconel 600 substrates. Inconel was chosen because it is characterized by high strength and environmental resistance at elevated temperatures [58]. Prior to Pyromark application, the Inconel substrates were degreased with 1–bromopropane and grit-blasted using 60–120 mesh garnet abrasives to increase surface roughness, which reduces the likelihood of delamination during curing [44]. Afterward, the substrates were cleaned again with 1–bromopropane. The samples underwent a final cleaning with methyl-ethyl ketone and were air-dried before painting began. Finally, the Pyromark was mixed for 20 min to eliminate sediment and ensure uniformity. This procedure was consistent with the application method outlined in [38].

Pyromark was applied to the Inconel substrates using a high-volume low-pressure spray gun at roughly 35 psi. The number of layers applied to each substrate was varied to analyze the impact of coating thickness on Pyromark degradation. For the samples with multiple layers, each layer dried at room temperature for approximately 10 min—until the paint was no longer visibly wet—before the subsequent layer was applied. The risk of delamination during the curing process was further diminished by air-drying the samples for at least 36 h [40]. The samples were cured in a Keith KSK-12 oven using the following procedure:

1) Heat to 120°C, hold for 2 h;

2) Heat to 250°C, hold for 2 h;

3) Heat to 400°C, hold for 2 h;

4) Heat to 540°C, hold for 1 h;

5) Heat to 800°C, hold for 1 h.

This curing procedure adheres to the method described in [38], but step 3 was added to further prevent delamination [59]. For the final step in the curing process, Ho *et al.* [38] recommend heating the Pyromark samples to 50°C above the maximum anticipated surface temperature, which was 750°C in this work. A heating rate of 5°C/min was used during the curing process, as Pyromark coatings are more likely to remain intact when cured at rates below 10°C/min [38], [44].

After curing was complete, the thickness of each sample was measured using a Mitutoyo CD-AX Series Absolute Digimatic caliper. Table 1 lists the number of layers applied to each substrate, the average total coat thickness, and the average thickness per layer for each sample. Due to asynchronous fabrication of the samples, the average thickness of each Pyromark layer was not uniform across the samples. These discrepancies were not concerning because Pyromark coatings on real central receivers vary in thickness from panel to panel [60], and the total coat thickness of the samples was consistent with prior studies on Pyromark application [40], [44], as well as Pyromark coatings on actual central receivers [9].

**Table 1.** Number of layers, average total coat thickness, and average thickness per layer of each sample.

| Sample | Number of layers | Average total coat thickness [µm] | Average thickness per layer [µm] |
|---|---|---|---|
| S1 | 1 | 12.5 | 12.5 |
| S2 | 2 | 27.5 | 13.8 |
| S3 | 3 | 70.0 | 23.3 |
| S4 | 4 | 77.5 | 19.4 |

The four samples were subjected to six oven treatments at 750°C. This temperature was chosen because it is consistent with the expected surface temperature of next-generation central receiver systems [10]. During each treatment, a Keith KSK-12 oven was heated to 750°C at a rate of 5°C/min, held at a specified dwell time, then cooled to 20°C at a rate of 5°C/min. The total aging duration was 151 h, or 18.9 days of CSP power plant operation, assuming a daily operation time of 8 h [45]. The oven treatments are summarized in Table 2.

**Table 2.** Time of each oven treatment and the total time of oven treatments.

| Treatment | Treatment duration [h] | Total time [h] |
|---|---|---|
| 1 | 1 | 1 |
| 2 | 10 | 11 |
| 3 | 20 | 31 |
| 4 | 30 | 61 |
| 5 | 40 | 101 |
| 6 | 50 | 151 |

## 2.2. Absorptivity measurements and results

A Hitachi U-4100 spectrophotometer with an integrating sphere detector was used to measure the spectral, hemispherical reflectivity $\rho_\lambda$ of the Pyromark samples at room temperature for wavelengths ranging from 200 nm to 2500 nm at an interval of 5 nm. Available data for the ASTM G-173 spectra [61] was utilized to calculate the samples' total, hemispherical absorptivity by:

$$\alpha = \frac{\int_0^\infty \alpha_\lambda G_\lambda d\lambda}{\int_0^\infty G_\lambda d\lambda} \tag{1}$$

where $\alpha_\lambda$ (= $1 - \rho_\lambda$) is the spectral, hemispherical absorptivity and $G_\lambda$ is the global spectral irradiance from [61].

Figure 1 shows the samples' total, hemispherical absorptivity as a function of aging time at 750°C. Prior to isothermal aging, the four samples' total, hemispherical absorptivity values were within 0.33% of each other. S1—the thinnest Pyromark coating—maintained the highest and most consistent total, hemispherical absorptivity of the four samples. The difference between S1's minimum and maximum total, hemispherical absorptivity throughout thermal aging treatments was within the spectrophotometer's photometric accuracy and could be attributed to instrument error. In [40], a Pyromark-coated Inconel 625 substrate with a comparable coating thickness to S1 experienced a much larger drop in total, hemispherical absorptivity in 100 h of aging at 750°C. Differences in the curing procedure and curing ramp rate may explain this discrepancy. Figure 1 shows that S2 and S3's total, hemispherical absorptivity declined more rapidly during earlier oven treatments, then stabilized. This aging pattern was also observed in [13], [21], [37], [38], [40], although the treatment duration before stabilization occurred in [21], [38], and [37] was much longer. S4—the thickest Pyromark coating—exhibited the largest decrease in total, hemispherical absorptivity, and its aging pattern deviated from those exhibited by the other samples. However,

S4's decrease in total, hemispherical absorptivity as a function of thermal aging was much closer to values in the literature [36], [38], [40].

S1's stability throughout thermal aging suggested thinner Pyromark coatings were advantageous. This conclusion was supported by S4's rapid degradation and significantly lower total, hemispherical absorptivity after 151 h of aging. However, S3 maintained a slightly higher total, hemispherical absorptivity than S2 throughout thermal aging despite its thicker coat.

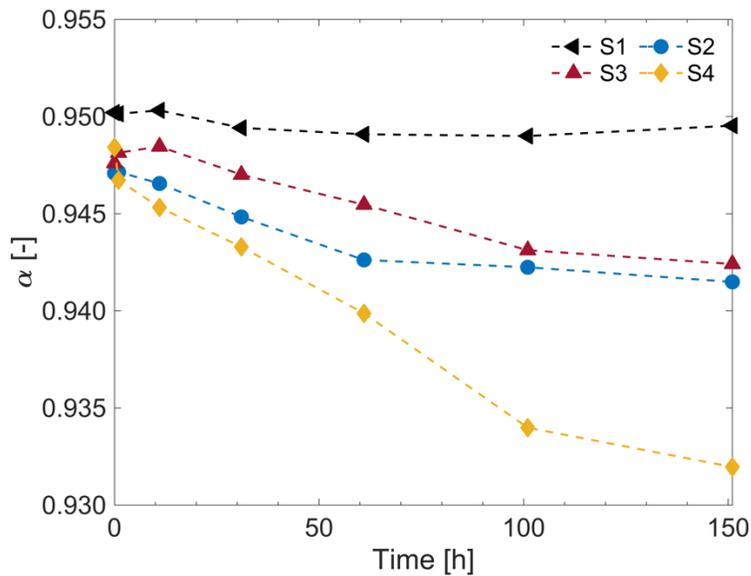

**Figure 1.** Samples' total, hemispherical absorptivity as a function of thermal aging at 750°C. The markers denote separate spectrophotometer measurements after each oven treatment.

Shifts in Pyromark's hue after aging have been reported previously [19], [21], [34], [37], [44], [46] and were expected in these experiments. Photos of the samples were taken before thermal treatments began and after each treatment to track evolutions in their color since radiation intensity of the solar spectrum is highest in the visible range [21]. Figure 2 depicts the samples in their initial state (no aging) and final state (151 h of thermal aging). After 151 h of aging, all four of the Pyromark coatings appeared to lighten from deep black to a muted black with brown and gray undertones, and certain areas of the coatings on each sample lightened more drastically after aging. This nonuniform change in color was attributed to variation in the coat thickness of individual samples, which has been observed in Pyromark coatings that were applied manually with a spray gun [9]. None of the samples exhibited signs of delamination during isothermal aging, and all four coatings laid completely flat on the substrate after 151 h at 750°C.

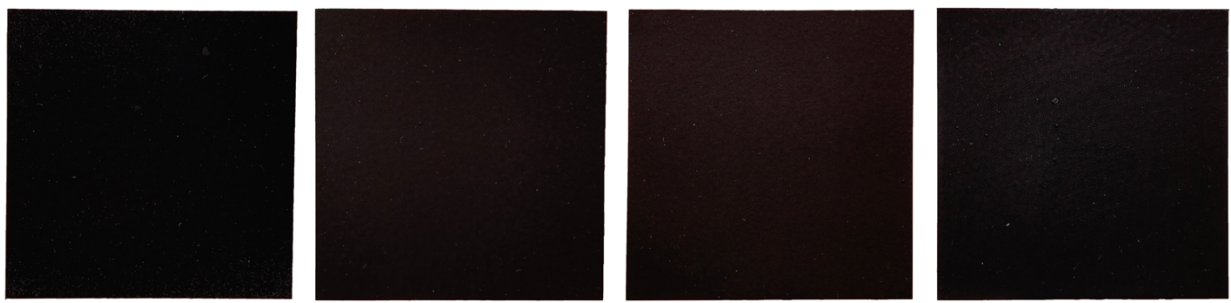

(a)

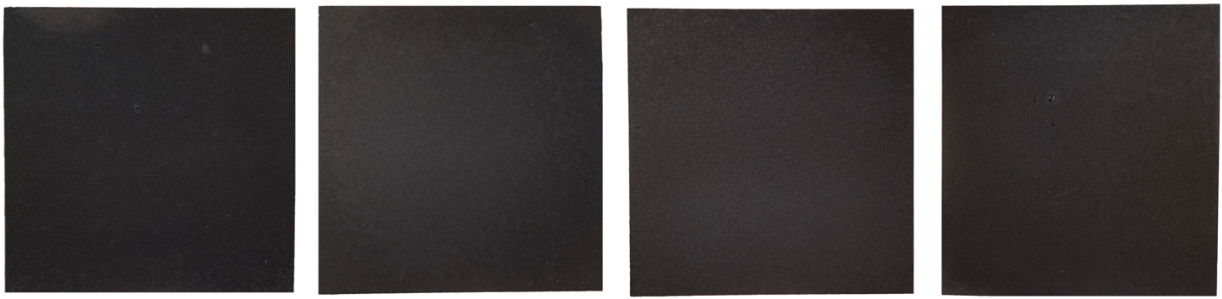

(b)

**Figure 2.** Photos of (left to right) S1, S2, S3, and S4. (a) Before isothermal aging treatments. (b) After 151 h of thermal aging at 750°C.

## 3. FOM and heat losses as a function of thermal aging

### 3.1. FOM and heat loss modeling

An FEA model was created in Abaqus [62] to quantify the impacts of Pyromark thermal aging on the heat losses experienced by a central receiver and to calculate an FOM that included all heat losses as a function of thermal aging. A typical central receiver consists of multiple panels in an approximate cylinder [50], [56]. For simplicity, the central receiver was modeled as a small area of a single receiver panel (3.81 cm × 3.81 cm × 0.05 cm plate with a uniform mesh of 9025 elements). The thickness of the Pyromark coating was also neglected because it was small compared to the substrate thickness. Convection, reflection, and emission losses were incorporated in the model's boundary conditions, as shown in Fig. 3.

The temperature distribution in the plate was calculated by solving the heat diffusion equation for three-dimensional, steady-state heat transfer without heat generation ($\nabla^2 T = 0$). The heat flux vector was then obtained from the temperature distribution via Fourier's law, $q'' = -k\nabla T$, where a thermal conductivity $k$ of 25.6765 W/(m·K) was used for Inconel 600 at 1000 K [63]. Solar radiation was modeled as a uniform heat flux acting normal to the front surface of the plate. Specifically, the absorbed solar flux $q''_{abs}$ represented the proportion of incident solar radiation absorbed by the receiver at the boundary $x = 0$:

$$q''_{abs} = \alpha C q''_{sol}. \tag{2}$$

In this model, $C$ was 1000, which is typical for central receiver systems [1], $q''_{sol}$ was 1000 W/m², and $\alpha$ was obtained from spectrophotometer measurements depicted in Fig. 1.

The height of the tower upon which central receivers sit and the elevated temperature of the receiver facilitates free and forced convection. Two scenarios were considered: 1) pure free convection on all surfaces of the plate in contact with the ambient air, which represented still wind conditions; and 2) free convection on the edges of the plate and forced convection on the plate's front surface, which represented windy conditions. Both convection scenarios are depicted in Fig. 3(a). Calculating an accurate forced convection heat transfer coefficient $h_{forced}$ introduced challenges beyond the scope of this work. Therefore, the forced convection heat transfer coefficient was defined as 50 W/(m²·K). Based on theoretical [50], [52] and measured [49] convection heat transfer coefficients for central receivers, this prescribed value of $h_{forced}$ was conservative. sCO₂ was chosen as the heat transfer fluid due to the substantial interest in sCO₂ Brayton cycles for next-generation CSP systems [10], [11]. The heat transfer coefficient of 18,000 W/(m²·K) and temperature of 1000 K for sCO₂ were based on findings in [64].

The radiation boundary conditions varied by surface, as shown in Fig. 3(a) and Fig. 3(b). Since the front of the plate was the only surface covered with Pyromark, the boundary $x = 0$ was defined by the total, hemispherical emissivity of Pyromark at 750°C, or 0.87 [38]. Lower emissivity values have been measured for Pyromark [22], but 0.87 was chosen as a conservative value. The model used the total, hemispherical emissivity of Inconel 600 on the top, bottom, and sides of the plate. At 750°C, Inconel's emissivity ranges from ~0.4 to ~0.9 depending on its oxidation state [65], which was unknown for these samples. A total, hemispherical emissivity of 0.9 was chosen as a conservative value.

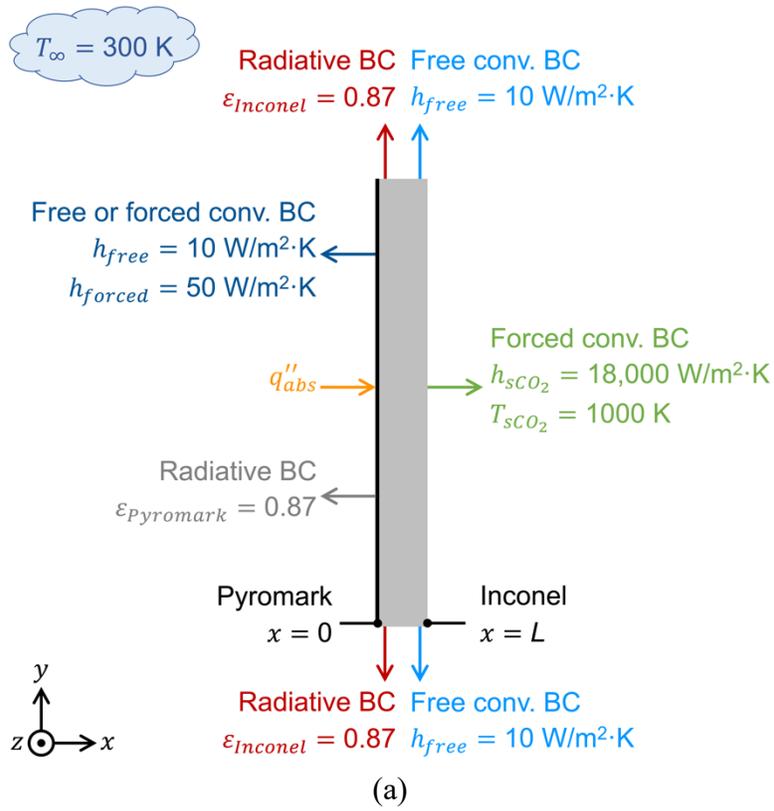

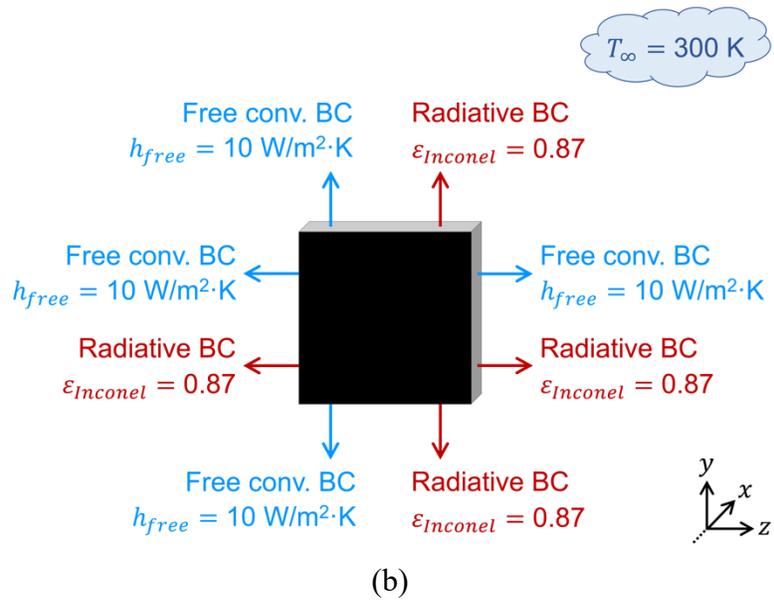

**Figure 3.** Boundary conditions (BCs) for the FEA model used to evaluate the heat losses and FOM. (a) Side view of the BCs on top, bottom, front and back of the plate. (b) Diagonal view of the BCs on the top, bottom, and sides of the plate.

Pyromark's degradation after thermal aging affects the reflection, emission, and convection losses experienced by a receiver. These heat losses then impact the FOM. Reflection losses were defined as the amount of incident solar energy not absorbed by the receiver:

$$q_{loss,refl} = A_{front} C q''_{sol}(1 - \alpha) \qquad (3)$$

where $A_{front}$ is the surface area of the front face of the plate.

Emission losses quantified the radiation emitted by the Pyromark coating as a function of temperature:

$$q_{loss,emiss} = A_{front} \varepsilon \sigma T^4 \qquad (4)$$

where $T$ is the maximum temperature of the plate's front surface, as determined by the FEA model.

Both free and forced convection losses from the front surface of the plate were modeled by Newton's law of cooling:

$$q_{loss,conv} = h A_{front}(T - T_\infty) \qquad (5)$$

where $T_\infty$ is the ambient temperature fixed at 300 K.

Figure 3(b) shows that radiation exchange between Inconel and the ambient governed emission losses on the top, bottom, and sides of the plate, while pure free convection governed the convection losses on these surfaces. Since the thickness of the samples was much smaller than their height and width, the emission and convection losses from the plate's front surface far outweighed the losses from the top, bottom, and sides. The emission and convection losses from these surfaces were merged and evaluated as "side losses":

$$q_{loss,side} = q_{loss,total} - q_{loss,refl} - q_{loss,emiss} - q_{loss,conv}. \qquad (6)$$

The variable $q_{loss,total}$ represents the total thermal losses experienced by the samples, or the difference between the incident solar irradiation at the boundary $x = 0$ and the heat flux transferred to the sCO₂ at the boundary $x = L$:

$$q_{loss,total} = A_{total}(Cq''_{sol} - q''_{x=L}). \tag{7}$$

In (7), $A_{total}$ is the total surface area over which losses occurred. This parameter excluded the area of the plate's back surface because no thermal losses were associated with convection heat transfer between the back of the plate and the sCO$_2$. The heat flux converted into thermal energy at the boundary $x = L$, $q''_{x=L}$, was output by the FEA model and included reflection, emission, and convection losses.

In this work, the typical FOM definition in the literature [14] discussed in Section 1 was classified as an idealistic scenario due to the omission of convection losses:

$$(FOM)_{ideal} = \frac{\alpha Cq''_{sol} - \varepsilon\sigma T^4}{Cq''_{sol}}. \tag{8}$$

As noted below (4), $T$ was determined by the FEA model, which included convection and radiative boundary conditions. Therefore, convection losses had an indirect impact on the ideal FOM through emission losses that depended on $T^4$.

The actual FOM was defined as the ratio of the heat flux transferred to the sCO$_2$ at the boundary $x = L$ to the total incident solar irradiation:

$$(FOM)_{actual} = \frac{q''_{x=L}}{Cq''_{sol}}. \tag{9}$$

Unlike the ideal FOM, the actual FOM explicitly accounted for all heat losses (i.e., reflection, emission, and convection) via $q''_{x=L}$.

### 3.2. FOM predictions as a function of thermal aging

The ideal and actual FOMs as a function of thermal aging are compared in the absence [Fig. 4(a)] and presence [Fig. 4(b)] of wind. For all samples, the ideal FOM was slightly larger when wind was flowing, which is misleading. The maximum temperature on the plate's front surface $T$ was lower in windy conditions than in still wind conditions owing to a larger heat transfer coefficient, and this consequently reduced the emission losses in (8). Conversely, the actual FOM was lower in windy conditions than still wind conditions, which was consistent with computed convection losses in the 10 MW$_e$ Solar Two central receiver [51].

In the absence of wind, the difference between the ideal and actual FOMs after 151 h of thermal aging was roughly 0.83% for all samples. Equation (8) therefore provided a decent approximation of Pyromark's FOM under still wind conditions, but such conditions are unlikely, as the wind velocity around real central receivers typically ranges from ~8 m/s to ~20 m/s [50], [51], [66], [67]. Thus, the ideal FOM is inaccurate for evaluating the performances of actual central receivers. When wind flow was considered, the difference between the samples' ideal and actual FOMs after 151 h was ~4.40%.

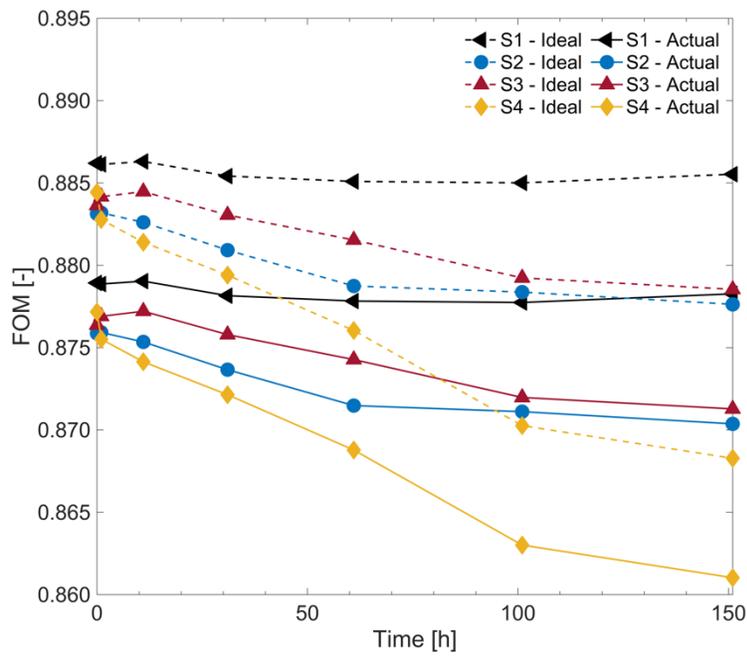

(a)

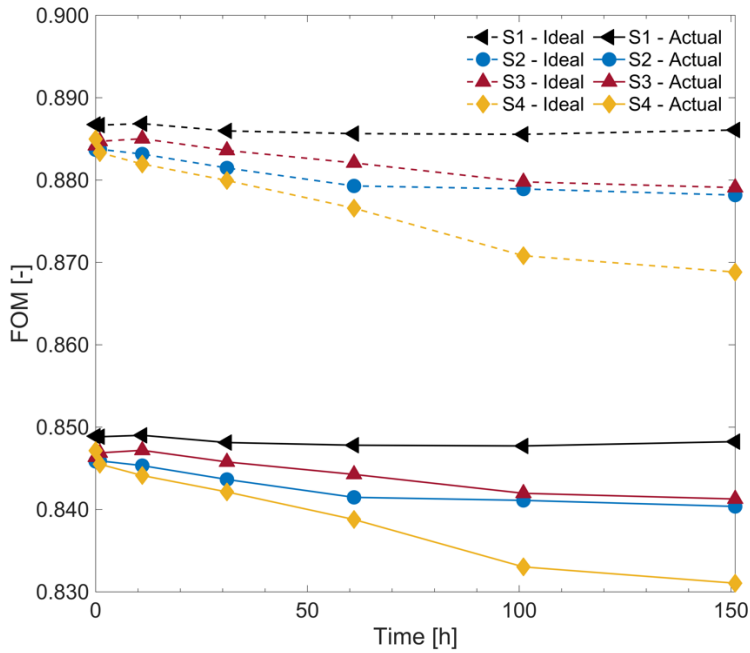

(b)

**Figure 4.** Ideal and actual FOMs as a function of thermal aging. (a) Free convection boundary condition, or still wind conditions. (b) Forced convection boundary condition, or windy conditions.

The curves in Fig. 1 and Fig. 4 are nearly identical, which confirmed the strong relationship between total, hemispherical absorptivity and FOM. This result was expected due to the high concentration ratio of central receiver systems. S1's FOM remained the most stable throughout thermal aging and was the highest of the four samples after 151 h at 750°C. Still, its actual FOM after 151 h was roughly 0.85 for the forced convection boundary condition. S4's decline in FOM was the most severe, but the actual FOMs of S2 and S3 still dropped 0.60% and 0.65%, respectively, under windy conditions. Considering the scale of central receiver systems, this decrease was substantial. Of the 12 projects involving solar power towers currently under construction globally, 10 have a nominal capacity greater than 100 MW [68]. Therefore, a 0.60% decrease in the receiver's FOM could abate power production on the order of kW.

In 2012, the U.S. Department of Energy launched the CSP SunShot initiative and set a target thermal efficiency of 0.90 for receiver subsystems [69]. The expression for thermal efficiency was equivalent to the ideal FOM given by (8). Figure 4 reveals that the samples did not achieve this goal efficiency at any point between 0 h and 151 h of aging for either convection boundary condition. After 151 h of aging, S1's total, hemispherical absorptivity was within 0.05% of the value reported by the manufacturer for pristine Pyromark [31], yet its ideal FOM fell short of the SunShot goal efficiency by 1.62% and 1.55% for the free and forced convection boundary conditions, respectively.

### 3.3. Analysis of heat losses as a function of thermal aging

Figure 5 depicts S3's distribution of heat losses before [Fig. 5(a)] and after [Fig. 5(b)] thermal aging for the forced convection boundary condition. S3 provided a good representation of the other samples' distribution of heat losses. Before thermal aging, all the samples' emission losses constituted the largest proportion of heat losses, which substantiated prior claims that Pyromark's main shortcoming is its high emissivity [2]. Generally, the samples' emission losses remained the largest thermal loss after 151 h of aging, but the percentage of total heat losses due to emission decreased as the percentage due to reflection losses increased, as shown in Fig. 5(b); this was caused by the samples' decrease in total, hemispherical absorptivity. Convection losses comprised nearly a quarter of the samples' total heat losses, indicating that the ideal FOM neglected a sizeable proportion of the samples' heat losses. Although the forced convection heat transfer coefficient in this work, 50 W/(m$^2$·K), was higher than published values for central receivers [49], [50], [52], the percentage of the samples' total heat losses attributed to convection losses agreed with other percentages found in the literature [7], [48], [52]. The side losses for each sample were only 5% of the total heat losses, hence combining free convection losses and radiative losses from the top, bottom, and sides of the samples was justified.

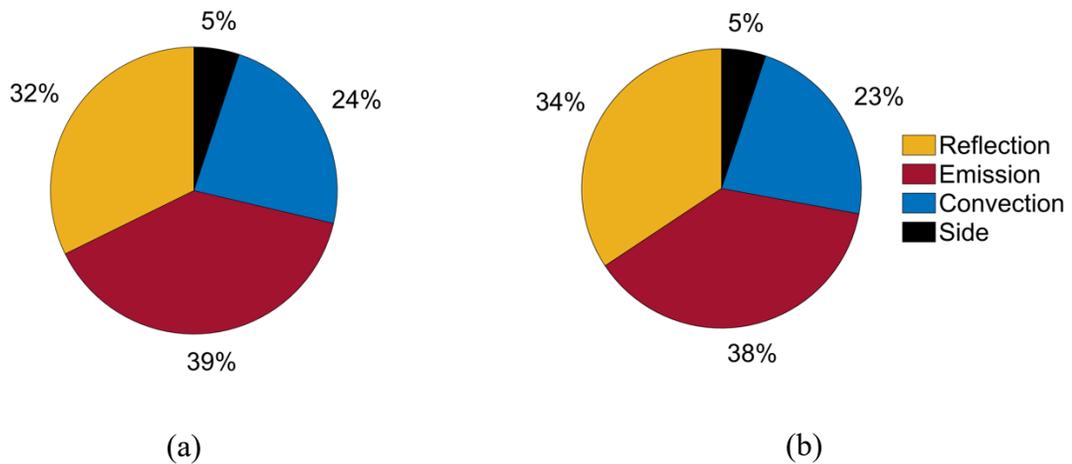

(a)          (b)

**Figure 5.** S3's proportion of reflection, emission, convection, and side losses for the forced convection boundary condition. (a) Before isothermal aging treatments. (b) After 151 h of isothermal aging at 750°C.

Figure 6 illustrates the heat loss distribution for S4 before [Fig. 6(a)] and after [Fig. 6(b)] thermal aging. The proportions of S4's total heat losses changed more drastically than the other samples' between 0 h and 151 h of aging due to its unstable total, hemispherical absorptivity. S4's heat loss distribution was nearly identical to the other samples before thermal aging treatments—specifically, emission losses dominated its total heat losses. After 151 h of aging, however, reflection losses constituted the highest percentage of total heat loss, as shown in Fig. 6(b). This shift in S4's heat loss percentages demonstrated that reflection losses can exceed emission losses if Pyromark's total, hemispherical absorptivity degrades significantly. This result was only applicable to CSP systems with large concentration ratios since Pyromark's decline in total, hemispherical absorptivity was intensified by the large value of $C$ per (3).

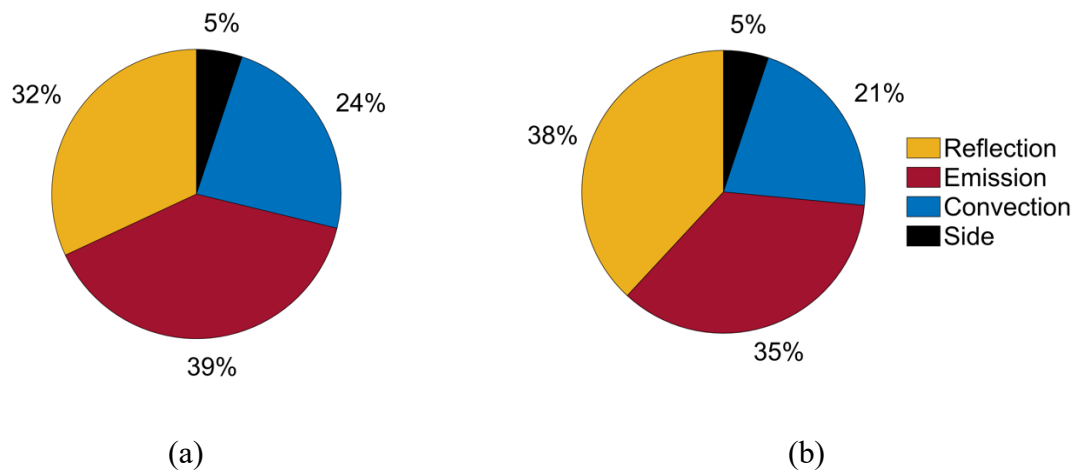

(a)          (b)

**Figure 6.** S4's proportion of reflection, emission, convection, and side losses for the forced convection boundary condition. (a) Before isothermal aging treatments. (b) After 151 h of isothermal aging at 750°C.

## 4. Conclusions

This work characterized Pyromark 2500's total, hemispherical absorptivity, FOM, and heat losses as a function of thermal aging. Assessing the heat losses associated with Pyromark degradation and calculating an FOM that accounted for reflection, emission, and convection losses were of particular interest. Four Pyromark-coated Inconel 600 substrates were subjected to 151 h of isothermal aging treatments at 750°C, and the samples' total, hemispherical absorptivity was recorded throughout aging. The total, hemispherical absorptivity values were input into an FEA model, and the outputs of this model were used to calculate each sample's reflection, emission, and convection losses, as well as FOM.

The four Pyromark coats were manufactured with various thicknesses. The sample with the thinnest Pyromark coat (S1) maintained the most stable total, hemispherical absorptivity throughout thermal aging, suggesting that thinner Pyromark coatings were advantageous. This inference was supported by the immediate degradation and poor total, hemispherical absorptivity of S4, which had the thickest Pyromark coat, after 151 h of aging at 750°C. However, samples with comparable coating thicknesses did not perform equally. Ultimately, no conclusions could be drawn about optimal coating thickness because the sample size in this work was too small for statistical analysis. Additional experiments with a minimum of 30 Pyromark samples, such as [40] and [44], would be necessary to make definitive conclusions about the relationship between Pyromark's coating thickness and total, hemispherical absorptivity.

An ideal FOM that applied the most common FOM equation in the literature was compared to an actual FOM equation that incorporated all major heat losses experienced by central receivers. After 151 h of thermal aging, the ideal FOM, which neglected convection losses, overestimated Pyromark's FOM by approximately 4.40%. Convection losses comprised between 21% and 24% of total heat losses, depending on the sample. Thus, neglecting the convection losses experienced by absorber coatings and central receivers ignores a considerable proportion of total heat losses. After S4's total, hemispherical absorptivity dropped during thermal aging, its reflection losses comprised a larger percentage of the sample's total heat losses than emission losses. Solar selective absorber coatings have been developed to mitigate emission losses in the infrared spectral band, but the proportions of S4's heat losses suggested that reflection losses will thwart a coating's performance more than emission losses if the coating's total, hemispherical absorptivity degrades considerably. However, this finding was only relevant for CSP systems with a large concentration ratio.

The isothermal aging treatments in this work were not as intensive or indicative of conditions around central receivers as other aging processes in the literature, such as cyclic aging or a combination of isothermal and cyclic aging referred to as cycle-and-hold [45]. These aging procedures have induced different aging patterns than isothermal aging [44], [45] and may result in different thermal loss distributions as well. Cyclic aging of Pyromark followed by an intensive heat loss analysis is recommended for future research. Investigations on Pyromark's performance after repainting are scarce in the literature [34]; thus, future experiments with aged Pyromark samples that have been repainted and aged again is also suggested.

**Acknowledgments**

M.F. acknowledges the support of the Natural Sciences and Engineering Research Council of Canada (NSERC) [funding reference number RGPIN-2023-03513]. K.B. acknowledges Dr. Joshua Fernquist and Dr. Stephen Naleway for their assistance with the isothermal aging treatments. She also acknowledges the academic support of Dr. Jacob Hochhalter and Dr. Sameer Rao throughout the course of these experiments.